\documentclass[12pt,preprint]{aastex}

\begin{document}

\title{The Diverse Stellar Populations of the W3 Star Forming Complex}

\author{Eric D.\ Feigelson and Leisa K.\ Townsley}

\affil{Department of Astronomy and Astrophysics, Pennsylvania State
University, 525 Davey Laboratory, University Park, PA 16802}

\shorttitle{Stellar populations of W3}

\shortauthors{Feigelson \& Townsley}

\email{edf@astro.psu.edu}

\slugcomment{To appear in ApJ}

\begin{abstract}

An $\sim$800~arcmin$^2$ mosaic image of the W3 star forming complex
obtained with the {\it Chandra X-ray Observatory} gives a valuable
new view of the spatial structure of its young stellar populations.
The {\it Chandra} image reveals $\sim 1300$ faint X-ray sources, 
most of which are PMS stars in the cloud.  Some, but not all, of the
high-mass stars producing hypercompact and ultracompact H{\sc II}
(UCHII) regions are also seen, as reported in a previous study.

The {\it Chandra} images reveal three dramatically different
embedded stellar populations.  The W3~Main cluster extends over 7~pc
with $\sim 900$ X-ray stars in a nearly-spherical distribution
centered on the well-studied UCHII regions and high-mass protostars.
The cluster surrounding the prototypical UCHII region W3(OH) shows a
much smaller ($\leq 0.6$~pc), asymmetrical, and clumpy distribution
of $\sim 50$ PMS stars. The massive star ionizing the W3~North H{\sc
II} region is completely isolated without any accompanying PMS
stars.  In W3~Main, the inferred ages of the widely distributed PMS
stars are significantly older than the inferred ages of the central
OB stars illuminating the UCHIIs.  We suggest that different
formation mechanisms are necessary to explain the diversity of
the W3 stellar populations: cluster-wide gravitational
collapse with delayed OB star formation in W3~Main, 
collect-and-collapse triggering by shock fronts in W3(OH), and a
runaway O star or isolated massive star formation in W3~North.

\end{abstract}


\section{Introduction \label{intro.sec}}

W3 (Westerhout 3) is perhaps the most active region of current star
formation in the nearby Galaxy.  Extending 30~pc along the edge of a
$M \simeq 5 \times 10^4$~M$_\odot$ giant molecular cloud (GMC), the
star-forming complex has dozens of embedded young massive stars
producing a variety of pre-stellar condensations, hot molecular
cores, hypercompact to small H{\sc II} regions, maser clusters, and
molecular outflows \citep[e.g.,][]{Lada78, Reid80, Dreher81,
Tieftrunk97, Chen06}. Its infrared sources have an integrated
luminosity of several~$\times~ 10^5$~L$_\odot$.  Situated just east
of W3 are the older IC~1795 and IC~1805 clusters, the latter lying
within the enormous W4 superbubble/chimney structure blown 
by generations of massive stars.  The W4--IC~1795--W3 complex 
is widely considered to be an examplar of sequential triggered
star formation  \citep{Lada78, Oey05}. Recent SCUBA observations
of the W3 GMC find a higher percentage of the gas mass gathered into
dense molecular clumps  at the eastern edge compared to the
undisturbed parts of the W3 GMC, supporting this triggering scenario
\citep{Moore07}. A detailed description of the W3 and W4 complexes
and a thorough review of the literature are given by \citet{Megeath07}.

The richest site of massive star formation in W3 is the W3~Main cluster
of embedded OB stars, dominated by the very young and luminous IRS~4 and
IRS~5 sources.  IRS~5 lies at the center of a 0.1~pc concentration of
massive stars resembling a nascent counterpart of the Orion Trapezium
\citep{Megeath05}.  W3(OH) to the southeast and W3~North to the north 
have massive stars but appear less active than W3~Main.  The distance to
the complex is accurately measured from maser kinematics to be 2.0~kpc
\citep{Xu06, Hachisuka06}.

Despite the intense study of W3 at radio, millimeter, and infrared
wavelengths, little is known about its low mass stellar population.
For example, a $JHK$ near-infrared (NIR) survey of $5\arcmin \times 5\arcmin$
in W3~Main reveals $\sim 40$ sources with $K$-band excesses indicative
of Class~I--II pre-main sequence (PMS) 
stars with disks \citep{Ojha04}.  Hundreds of other
stars are detected, but infrared photometry cannot discriminate disk-free
Class~III PMS stars from the strongly contaminating population of
unrelated Galactic field stars (mostly red giants).  A new mid-infrared
(MIR) photometric survey of W3~Main, IC~1795, and W3(OH) with the {\it
Spitzer Space Telescope} helps to identify cluster members \citep{Ruch07},
but it suffers confusion from three effects: foreground and background
Galactic field stars \citep{Benjamin05}, bright diffuse emission
produced by heated dust around the H{\sc II} regions \citep{Povich07}, 
and extragalactic objects with MIR excesses \citep{Harvey07}.

X-ray surveys of young stellar clusters (YSCs) with the {\it Chandra
X-ray Observatory} are surprisingly efficient at detecting low mass
PMS populations, even at distances around 2~kpc and at obscurations
$10<A_V<150$ mag typical for W3 stars.  PMS X-ray emission arises primarily
from violent magnetic reconnection events, similar to solar flares
but far more powerful, and is largely independent of circumstellar
disks or accretion \citep[see reviews by][]{Feigelson99, Feigelson07}.
Luminous and spectrally hard X-ray flares are present throughout the
PMS phases of Class~I--II--III at levels $10^2-10^3$ above that seen
in old disk populations \citep{Preibisch05a}, so relatively few Galactic
disk interlopers appear in X-ray samples.  These field star X-ray
sources and extragalactic contaminants are easily removed
\citep{Getman06}.  Due to a
poorly-understood statistical association between X-ray luminosity and
PMS stellar mass \citep{Preibisch05b,Guedel07}, a flux-limited X-ray
observation of a young stellar cluster will be roughly complete down to
a corresponding mass limit.

Taking together these properties of X-ray studies, we find that
X-ray surveys at sufficiently high spatial resolution and
sensitivity provide uniquely rich, largely disk-unbiased,
mass-limited, and nearly uncontaminated samples of PMS stars in both
embedded and unobscured YSCs. These samples complement MIR surveys
obtained with the {\it Spitzer Space Telescope}, which generally
extend down to lower masses (including the brown dwarf regime) but
cannot readily discriminate disk-free PMS stars from field stars.
{\it Spitzer} thus detects more disky Class~0-I-II systems while
{\it Chandra} effectively samples Class~III systems in addition to
many Class I and II stars.  The X-ray samples are
useful for various astrophysical purposes such as probing the
stellar Initial Mass Function, protoplanetary disk evolution, and
magnetic activity.

\begin{figure}
\centering
\includegraphics[height=0.8\textheight]{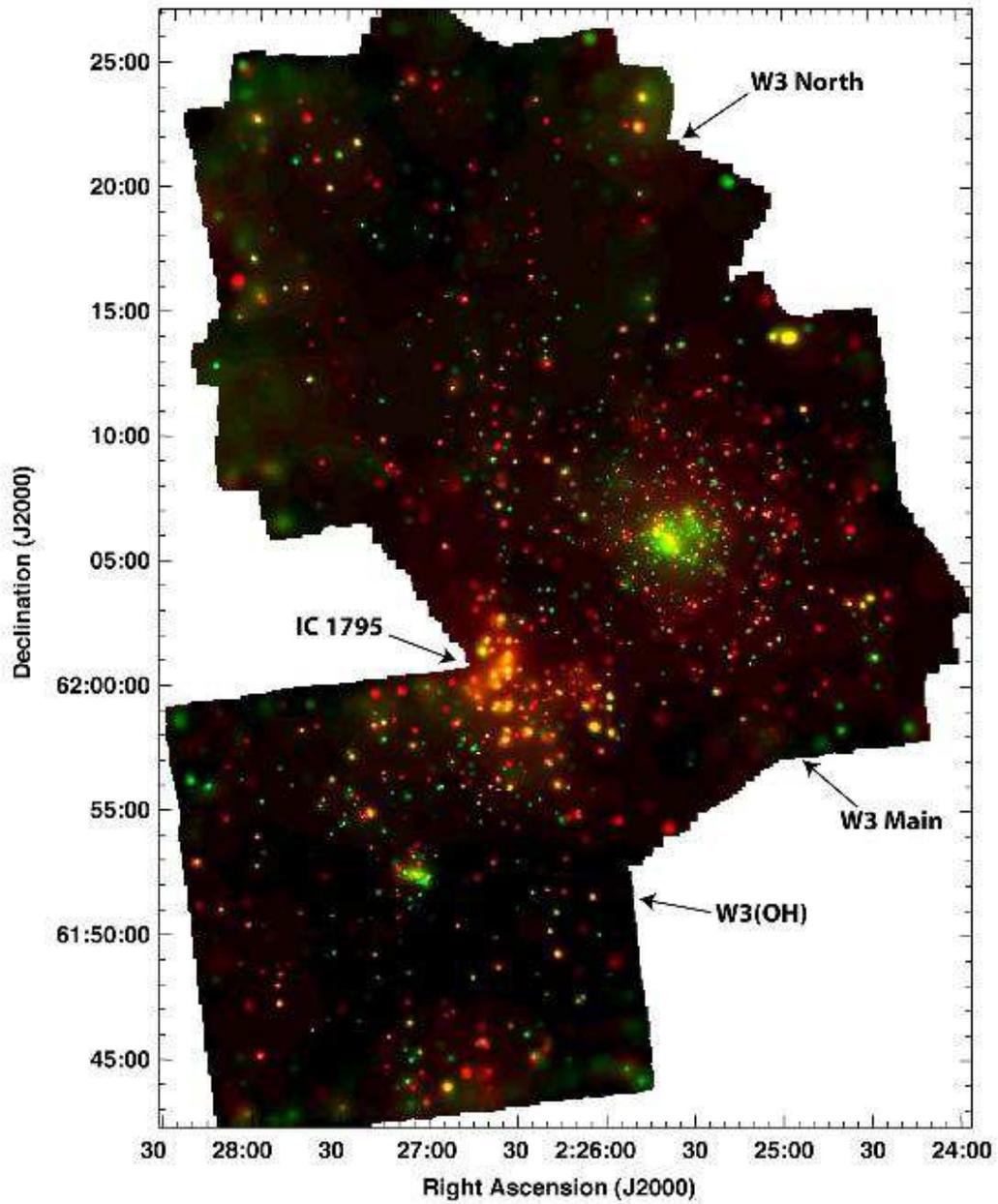}
\caption{A smoothed X-ray mosaic of W3 from the {\it Chandra}/ACIS
data where red intensity is scaled to the 0.5--2~keV emission and
green is scaled to 2--7~keV emission. The $\sim$800~arcmin$^2$
mosaic includes W3~North, W3~Main, part of the older cluster
IC~1795, W3(OH), and two small IR clusters located northeast of
W3(OH). Each of the seven ACIS-I pointings covers $17\arcmin \times
17\arcmin$. \label{fig:ACIS_mosaic}}
\end{figure}

In an early {\it Chandra} study, two $\sim$20~ks exposures of a
$\sim 300$~arcmin$^2$ field in W3~Main revealed 236 X-ray sources
\citep{Hofner02}.  Several are associated with massive stars ionizing
H{\sc II} regions but most do not have counterparts in $JHK$ images.
We report here an extension of those efforts with a {\it Chandra}
mosaic of 7 exposures totaling $\sim 230$~ks over $\sim 800$~arcmin$^2$,
spanning much of the W3 star forming complex 
(Figure~\ref{fig:ACIS_mosaic}).  A preliminary discussion of this mosaic, 
a {\it Chandra} Large Project, is given by \citet{Townsley06}.  Over 1300
X-ray sources are seen; a full listing and study of their properties
will be presented in a separate paper.  For each source, {\it Chandra}
observations provide a sub-arcsecond position, line-of-sight absorption,
and rough mass estimate in addition to magnetic activity characteristics.

We discuss here insights into the global structure and origins of
the W3 stellar populations derived from the new {\it Chandra} data.
The brief presentation of the observations in \S\ref{sec:obs} will
be expanded in a forthcoming paper with complete source lists similar to
our group's recent studies of the Cep~OB3b \citep{Getman06}, Pismis~24
\citep{Wang07a}, M~17 \citep{Broos07}, RCW~49 \citep{Tsujimoto07}, and 
Rosette star forming region \citep{Wang07b, Wang07c} YSCs.  
The three well-studied star forming regions in W3 are described and 
contrasted in \S\ref{sec:Xray}, and explanations for their origin are considered in
\S\ref{sec:diversity}.  Section~\ref{sec:W3Main.origin} considers in
more detail the implications of the W3~Main results for astrophysical
models of star cluster formation.

\section{Observations and data reduction \label{sec:obs}}

The X-ray observations were made with the Advanced CCD Imaging
Spectrometer (ACIS) camera on board the {\it Chandra X-ray Observatory}
\citep{Weisskopf02}.  Three contiguous regions of the W3 star forming
complex were observed with the $17\arcmin \times 17\arcmin$ ACIS imaging
array (ACIS-I) for roughly 80~ks each, divided into seven exposures:
three on W3~Main, one on W3(OH), and three on W3~North.  Except for
the two $\sim$20-ks W3~Main exposures discussed by \citet{Hofner02},
all were obtained between January and November 2005.

Data analysis followed procedures described in our group's previous ACIS
studies of young stellar clusters \citep[e.g.,][]{Townsley03,
Getman05, Wang07a, Broos07}.  X-ray events were corrected for CCD
charge transfer inefficiency \citep{Townsley02} and the data were
cleaned in a variety of ways \citep{Townsley03}.  The event data
were corrected to the {\it Hipparcos} reference frame by alignment
of bright on-axis X-ray sources with 2MASS stars then registered to
a common astrometric reference frame based on {\it Chandra}
boresights.  A preliminary source list was identified from the
merged observations using a wavelet-based algorithm
\citep{Freeman02}.  Individual positions are generally accurate to
$\pm 0.4$\arcsec\/ and double sources can be resolved at separations
$\ga 0.7$\arcsec. Images were created in soft (0.5--2~keV) and hard
(2--7~keV) bands and were corrected for exposure variations, then
adaptively smoothed using the CIAO tool {\it csmooth}
\citep{Ebeling06} to make the mosaic shown in
Figure~\ref{fig:ACIS_mosaic}.

\section{The X-ray stellar populations in W3  \label{sec:Xray}}

\subsection{W3 Main}

W3~Main is a massive YSC, famous for containing every known type of
radio H{\sc II} region from hypercompact to diffuse, 0.01 to 1~pc in
diameter, with ages $10^3$--$10^6$~yrs \citep{Tieftrunk97}.  These
H{\sc II} regions are embedded in a complex, highly clumped
molecular environment, with the younger (smaller) regions associated
with the densest clumps \citep[][and references therein]{Megeath07}.
Most radio and NIR studies have concentrated on the dense central
regions of the cluster, hence the complex was described as only
$\sim$4$\arcmin$ in size \citep{Tieftrunk97}.

Figure~\ref{fig:W3Main_small} shows this central region of the W3~Main
cluster in the mosaicked {\it Chandra} image.  X-ray point sources in our
preliminary source list are marked with blue circles; additional faint
X-ray sources are likely to emerge in our complete analysis, which will
involve image reconstructions of crowded regions \citep{Townsley06b}.
The radio H{\sc II} regions detailed in \citet{Tieftrunk97} are shown
schematically as magenta ovals.  X-ray sources that match the bright
infrared sources from Table~1 of \citet{Ojha04} are marked with green
circles and labeled.  Six of these sources were found in the earlier
{\it Chandra} study of this region \citep{Hofner02}.  Two sources in
the IRS5 region match NIR sources in Table~1 of \citet{Megeath05} and
are labeled using their nomenclature.

The wide-field {\it Chandra} mosaic (Figure~\ref{fig:ACIS_mosaic})
shows that the W3~Main YSC extends well beyond this central region;
it is rich and roughly spherical, resembling the clusters that
dominate many massive star forming regions.  Its full extent is
quite large; over 900 X-ray sources are distributed over the
17\arcmin\/ (10~pc) ACIS-I field centered around
($\alpha,\delta$)=(02:25:41,+62:05.9) close to W3~IRS5.  The cluster
is so large that for some stars, it is difficult to distinguish
membership in W3~Main from membership in the less absorbed IC~1795
cluster to the southeast.  Figure~\ref{fig:W3Main_smooth} shows how
the dense concentration of lower mass stars in the inner $\sim
1\arcmin$\/ centered around IRS~5, known from earlier NIR
observations \citep{Megeath96,Tieftrunk98,Ojha04}, extends smoothly
with the stellar surface density decreasing a factor of $\sim 300$
out to a radius of 5\arcmin\/ (3~pc). When the X-ray luminosity
function is corrected for contaminating sources and limited
sensitivity, and then scaled to the well-characterized Orion Nebula
Cluster \citep[e.g.][]{Wang07a}, the inferred cluster population
will be several thousand stars.  The stellar distribution in the
central region is nearly but not entirely symmetrical; an excess of
stars is seen $\sim 1$\arcmin\/ NW of IRS~5 around the W3~D and W3~H
UCHIIs compared to a symmetrical region SE of IRS~5 where no massive
stars are present.

The finding we emphasize here is not the previously known
concentration of high mass stars in the cluster core but the
richness, extent, and symmetrical appearance of the W3~Main stellar
cluster on scales of several parsecs.  The vast majority of these
sources are low-mass PMS stars with only minor contamination by
Galactic field stars or extragalactic sources
\citep[e.g.][]{Wang07a}. While the stellar concentration in the
central portion can be seen in $K$-band images
\citep{Megeath96,Tieftrunk98,Ojha04}, the full extent and simple
structure of W3~Main cannot be discerned in the NIR due to the
combination of patchy obscuration, nebular emission, and Galactic
field star contamination.

\subsection{W3(OH)}

W3(OH) is a rapidly developing UCHII region, seen in the radio as an
expanding shell of dense ionized gas around a heavily obscured ($A_V
\sim 50$) late-O star \citep{Dreher81, Turner84}. The expansion age
is $\sim 2 \times 10^3$~yr although an astrochemical model of the
molecular species suggests an age around $10^4-10^5$~yr
\citep{Kawamura98, Kim06}. Six arcseconds east of W3(OH) lies the
molecular hot core W3(H$_2$O) with three radio continuum peaks,
maser emission, and an unusual radio synchrotron jet \citep[][and
references therein]{Wyrowski99}. This small complex is sometimes
called the Turner-Welch Object. In a NIR image with $K<17.5$, a
stellar cluster with $\sim 200$ stars is seen with an elongated
$\sim 1\arcmin$\/ distribution around the UCHII region; this is the
one of the richest groupings of NIR stars within the huge W3/W4/W5
star forming complex \citep{Tieftrunk98, Carpenter00}. NIR data also
reveal two smaller clusters northeast of W3(OH) \citep{Tieftrunk98}.

\begin{figure}
\centering
\includegraphics[height=0.8\textheight]{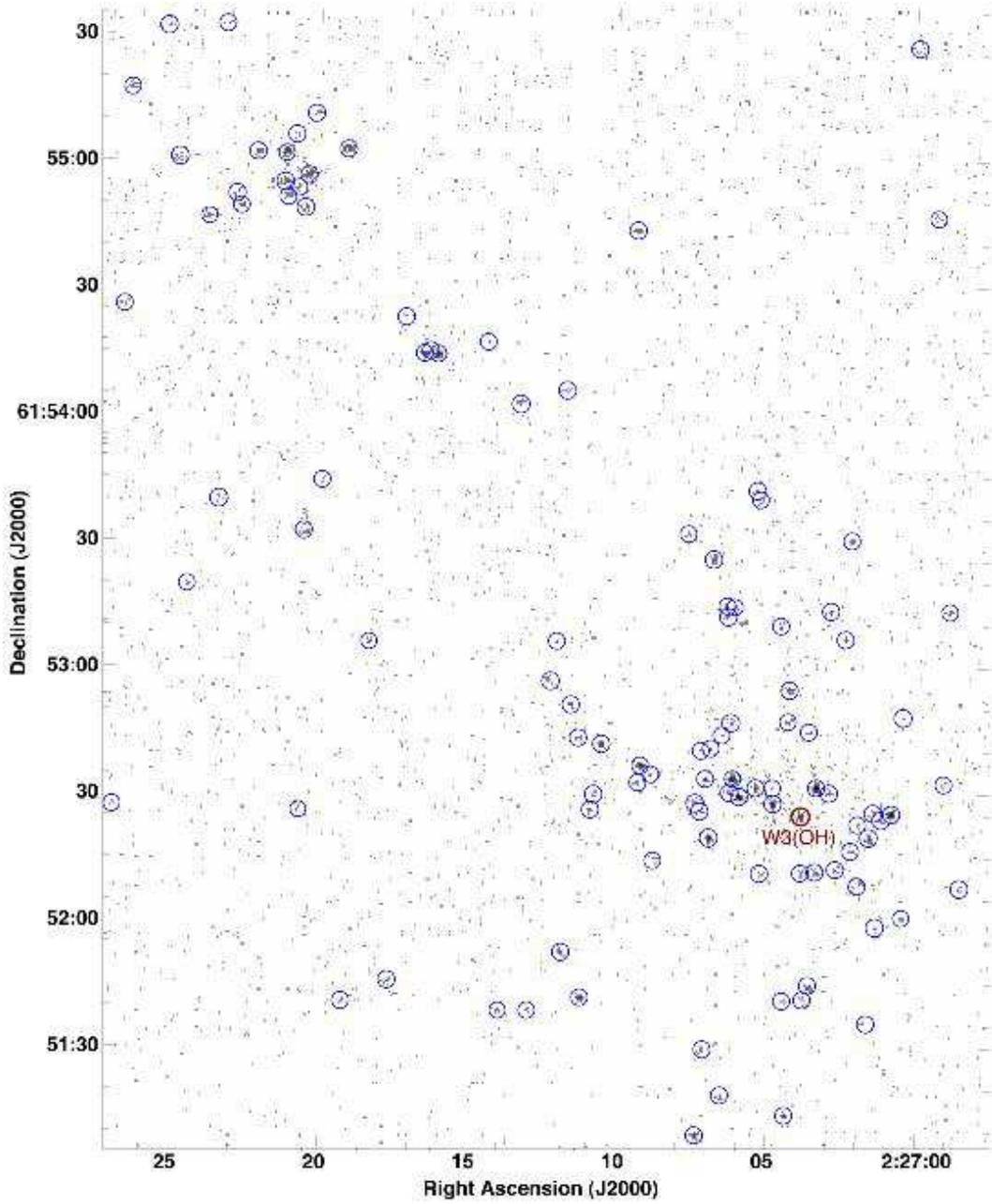}
\caption{ACIS image of the W3(OH) cluster and two small
IR clusters to its northeast ($3.5\arcmin \times 4.5\arcmin$, $2.0 \times 
2.6$ pc).  The X-ray source at the center of the UCHII region W3(OH)
is marked. \label{fig:W3OH}}
\end{figure}

The {\it Chandra} image shows that the YSC surrounding W3(OH)
is much smaller, sparser, and less symmetrical than W3~Main
(Figure~\ref{fig:W3OH}).  About 50 absorbed X-ray stars lie in a
region $0.5 \times 1$\arcmin\/ ($0.3 \times 0.6$~pc) oriented NE--SW
around W3(OH).  This cluster is accompanied by two sparse clumps with
about 5 and 20 stars, respectively, lying $\sim 0.5-1.5$~pc to the NE
of W3(OH), cospatial with small clusters seen in earlier NIR studies
\citep{Tieftrunk98}.

The young massive star ionizing W3(OH) is clearly detected in our
{\it Chandra} observation, at (02:27:03.84, +61:52:24.9).  It is a
surprisingly hard X-ray source; this hard emission allows it to be seen
through a large absorbing column ($A_V \sim 75$ mag) inferred from the
soft X-ray absorption.  The nearby high-mass system W3(H$_2$O), likely
powered by a protobinary of early-B stars \citep{Chen06}, is undetected
in our X-ray data.

\subsection{W3 North}

The bright H{\sc II} region G133.8+1.4 = W3~North is less
well-studied than the regions considered above.  The nebula is
excited by an optically visible O6 star, \#102 in the study of
IC~1795 by \citet{Ogura76} and \#7044 in the study by \citet{Oey05}.
It lies in a molecular cloud environment with density $\sim
600$~cm$^{-3}$ and mass $\sim 230$~M$_\odot$, more than an order of
magnitude below values for W3~Main and W3(OH) \citep{Thronson84,
Thronson85}. The ionized nebula is as bright as W3~A near IRS~5 in
W3~Main; it has a diameter of 2\arcmin\/ with estimated age $\sim
10^5$~yr \citep{vanderWerf90}.

\begin{figure}
\centering
\includegraphics[width=0.70\textwidth]{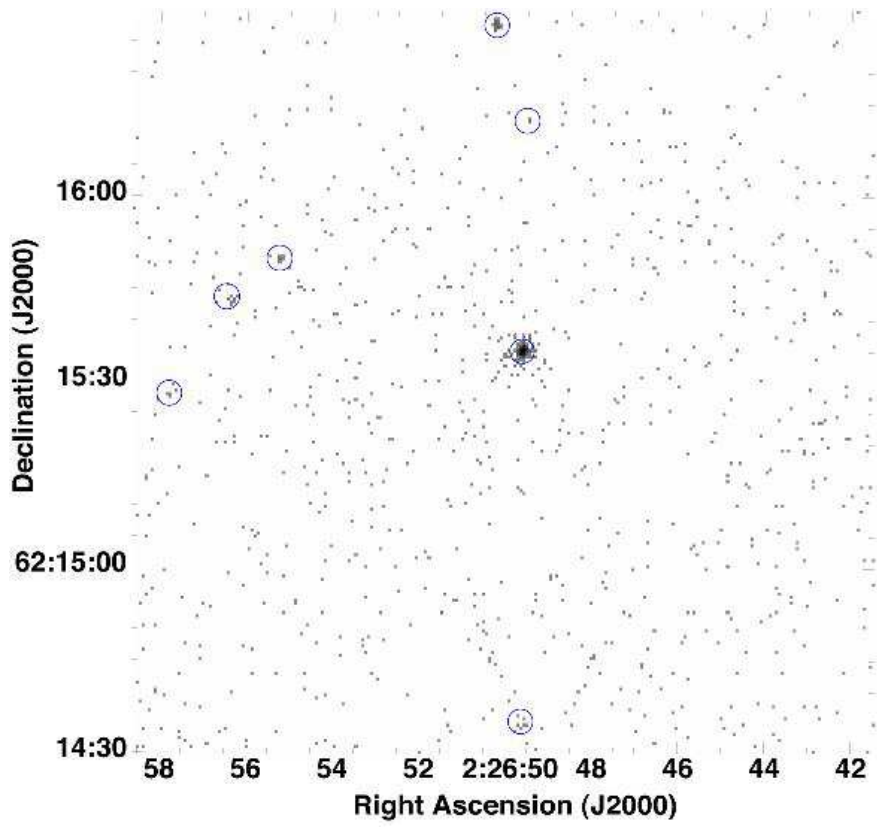}
\caption{{\it Chandra} image of W3~North ($2\arcmin$, 1.2 pc).  The
bright X-ray source at field center is the isolated O6 star. \label{fig:W3North}}
\end{figure}

We clearly detect the O6 star in our {\it Chandra} observation, at
(02:26:49.62, +62:15:35.0).  However the {\it Chandra} source
distribution around this O star differs dramatically from that seen
in either W3~Main or W3(OH):  no cluster of PMS stars is found in
its vicinity (Figure~\ref{fig:W3North}).  The nearest X-ray source
is $>35$\arcsec\/ distant and the local source density is consistent
with the general level of distributed young stars and contaminants
seen $5\arcmin-10$\arcmin\/ away.  The absence of a cluster in
W3~North was suggested by \citet{Carpenter00} based on NIR imagery,
and we strongly confirm this result with our X-ray observations.

\section{Interpreting population differences \label{sec:diversity}}

We find that three famous massive star-forming sites in the W3 cloud
show remarkable variety in their low mass stellar distributions: a
rich spherical cluster, an elongated collection of sparse star
clumps, and a completely isolated O star.  Evidence for these
differences was provided by earlier infrared studies, but the {\it
Chandra} dataset gives a more definitive view of this morphological
diversity from its more complete and unbiased sample of the low mass
population.  We can discuss the origin of the structures in W3.

The W3(OH) cluster is an order of magnitude smaller and roughly 20
times less rich than the W3~Main cluster.  The patchy distribution
of stars, elongated along an axis perpendicular to a vector pointing
towards the older IC~1795 cluster, supports a triggered origin due to
IC~1795 ionization and wind shock fronts as discussed by
\citet{Oey05}. The two small clusters, seen both in NIR \citep{Tieftrunk98} 
and X-ray images, lie along this same line and could have been
triggered by the same shocks. The morphology of the PMS
stellar distribution around W3(OH) resembles those seen in small
cometary globules \citep{Sugitani95, Getman07} and in larger
molecular clouds \citep{Zavagno06, Deharveng06, Broos07} at the
edges of H{\sc II} regions.  The elongation in stars around W3(OH)
appears perpendicular to the axis pointing towards IC~1795, similar 
to the elongation of the M17~SW stellar distribution which lies 
along the photodissociation region and perpendicular to the axis 
pointing towards the M17 central cluster NGC~6618 \citep{Broos07}. 
The W3(OH) structure has the fragmented and elongated 
appearance expected from the ``collect and collapse''
scenario of triggered star formation at the edge of an H{\sc II}
region \citep{Elmegreen77, Whitworth94, Dale07}.

For W3~North, we have a clear demonstration that its ionizing O star
is isolated, unaccompanied by a cluster of lower mass stars. The
simplest explanation is a runaway O star ejected from a rich
cluster in the W3/W4 region.  The W3~North radio continuum structure
does have a cometary tail on the SSE side of the ionizing star,
suggesting a northwesterly motion through the molecular medium
\citep{vanderWerf90}. This is inconsistent with an origin in W3~Main
which would require a northeasterly motion, but may indicate an
origin in the older IC~1795 or IC~1805 clusters. Accurate proper
motions are needed to test this model\footnote{The NOMAD catalog
\citep{Zacharias05} reports very large proper motions
($\sim$200~mas~yr$^{-1}$) for this star based on photographic sky
survey plates, but examination of the Digitized Sky Survey using
NASA's {\em SkyView} service shows a bright source in the same
location in both DSS1 and DSS2, implying that the NOMAD proper
motions are erroneous.  The region has bright irregular nebular 
emission, so mistakes can easily be made.}.

An alternative explanation, which seems feasible although
improbable, is an origin within the local W3~North molecular cloud.
Statistical simulations of sparse clusters with random draws from a
standard Initial Mass Function show a wide dispersion of maximum
stellar masses \citep{Bonnell99} and a few cases of field O stars
support occasional formation of massive stars in isolation
\citep{deWit05}.  In particular, the late-O star ionizing KR~140,
lying a degree south of W3~Main in the W3 molecular cloud, may have
formed in isolation \citep{Ballantyne00}.

Our findings do not support a simple unified origin of W3~North,
W3~Main, and W3(OH) as proposed by \citet{Oey05}.  In their
interpretation, the three regions of high-mass star formation are
components of a shell of molecular cloud material triggered into
gravitational collapse by the ionization and wind shocks produced by
the older IC~1795 star cluster lying east of the W3 molecular cloud.
In contrast, we find that only the W3(OH) stellar population has the
morphology expected from direct triggering by IC~1795 shock fronts.
The other two W3 stellar populations have very different morphologies:
the single O star
ionizing W3~North either formed in isolation or was dynamically
ejected from one of the richer nearby clusters, and the W3~Main
cluster has a spherical, centrally condensed appearance that does
not reflect the recent passage of a shock.

\section{The origin of the W3~Main cluster  \label{sec:W3Main.origin}}

\subsection{Two critical properties of W3~Main}

\begin{figure}
\centering
\includegraphics[width=0.40\textwidth]{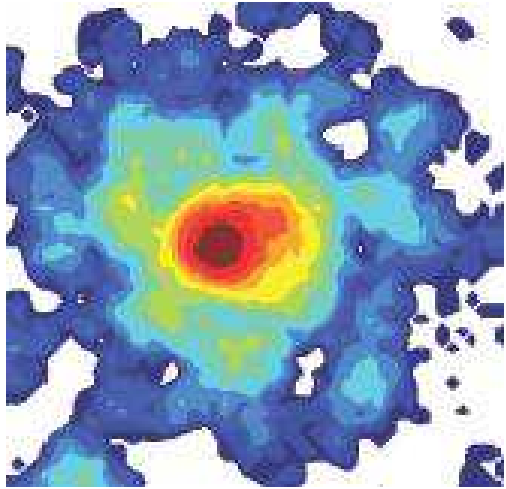}
\caption{Large-scale distribution of X-ray sources in W3~Main
(10\arcmin, 5.8~pc),
smoothed with a 30\arcsec\/ (0.3~pc) kernel.  Colors represent
logarithmic range of source surface density from 0.03 (dark blue) to
10 (brown) stars arcmin$^{-2}$. \label{fig:W3Main_smooth}}
\end{figure}

Two inferences can be made from the morphology shown in
Figure~\ref{fig:W3Main_smooth}.  These provide strong constraints on
the formation process of the rich W3~Main cluster.

First, as outlined above and shown in Figure~\ref{fig:W3Main_smooth}, 
the large-scale sphericity of the cluster
implies that the role of triggering by shocks from the older IC~1795
cluster (or by the W4 superbubble further to the east) discussed by
\citet{Oey05} is negligible, or at most indirect, in the sense that the 
star formation did not follow the passage of a localized shock.  
There is no elongation of the stellar distribution
along an East-West axis associated with a shock.  W3~Main was either
formed independently of an external trigger, or has dynamically
evolved so that evidence of its triggered origin has been erased.
The centrally concentrated, spherical morphology resembles the
distribution of X-ray stars in the Orion Nebula Cluster ionizing the
Orion Nebula \citep{Feigelson05}, the NGC~6618 cluster ionizing the
M~17 H{\sc II} region \citep{Broos07}, the NGC~2244 cluster ionizing
the Rosette Nebula \citep{Wang07b}, and many other YSCs. These stand
in contrast to the unconcentrated and elongated stellar distributions
attributable to shock triggering in small cometary globules
\citep{Sugitani95, Getman07, Ogura07} and in larger molecular clouds
\citep{Zavagno06, Deharveng06, Broos07} at the edges of H{\sc II}
regions.

The second inference concerning the origin of W3~Main to be made
from the {\it Chandra} image is that at least some of the OB
stars---those ionizing the well-studied hypercompact and
ultracompact H{\sc II} regions at the core of W3~Main---formed after
the bulk of the more widely distributed cluster PMS stars.  These
H{\sc II} regions have dynamical ages of $10^3-10^5$~yr
\citep{Tieftrunk97}\footnote{Under special circumstances, an H{\sc
II} region can appear as an UCHII at later times \citep{Franco07}.
This requires either that the O star is nearly stationary
($<<1$~km~s$^{-1}$ motion) at the center of a dense molecular core,
or that it has entered a second core at a later time. It seems
doubtful that this would fortuitously occur for several O stars in the
central region of W3, and in any case can not explain the
hypercompact regions in W3~M.}. If the PMS stars had similar ages,
they would all be Class~0--I protostars.

However, only a few percent of ACIS sources are Class~II, and $<1$\%
appear to be Class~O--I in the \citet{Ruch07} dataset of the
brighter {\it Spitzer} sources.  It is not possible that there
exists a vast population of Class~O--I sources undetected by {\it
Spitzer} in the outer regions of W3~Main.  In the central region
around IRS~5, whre {\it Spitzer} sensitivity is limited by the bright
diffuse emission, \citet{Megeath96} found that no more than 30\% of
the NIR sources were Class~I.  This implies that most of the PMS stars
in W3~Main are Class~III, as in most other young stellar clusters
observed with {\it Chandra}, and the age of the low-mass population
is $> 0.5$~Myr.  Even in clusters rich in Class 0 protostars, such
as NGC~1333, many X-ray sources are Class II and III
systems \citep{Getman02}.  This discrepancy in W3~Main
may constitute the best case that a PMS population is much older ($>
0.5$~Myr) than at least part of its associated OB population ($<
0.1$~Myr).

If the PMS stars are characteristically $10^6$ years old and some of
the central OB stars are $< 10^5$ years old, those OB stars must
have formed after the lower mass stars. This form of age spread has
long been noted in older stellar clusters from studies of HR
diagrams \citep{Herbst82, Adams83, Doom85, Shull95, DeGioia01}.  A
young age has also been indirectly suggested for the Trapezium OB
stars in Orion based on the inferred short lifetimes of proplyds in
the presence of ultraviolet photoevaporation \citep{ODell98}.  The 
W3~Main OB stars are directly confirmed to be extremely young and 
still forming based on their  very small H{\sc II} regions; this is crucial
for establishing that the central OB stars formed after the larger PMS
population.  

\subsection{Implications for the formation of W3~Main}

Together, these two results strongly preclude the application of an
old and simple model of cluster and high mass star formation (see 
reviews by Bonnell et al.\ 2007 and Larson 2007). Pre-stellar molecular
cloud condensations were traditionally thought to be centrally
concentrated with higher densities $\rho$ at the center; e.g., an
isothermal equilibrium Bonnor-Ebert sphere.  The free-fall time is
then shorter at the core, $t_{ff} \propto \rho^{-1/2}$, implying
rapid gravitational collapse and fragmentation.  Gas quickly falls
into the central region where, if Bondi-Hoyle accretion is
unimpeded, the more massive protostars tend to grow fastest
according to $\dot{M} \propto M^2$.  Disk accretion of high mass
protostars can be very rapid with $\dot{M} \sim
10^{-4}$~M$_\odot$~yr$^{-1}$ implying full growth in $\sim 10^5$~yr
\citep[see review by][]{Cesaroni07}.  In these simple scenarios for
cluster formation, OB stars concentrated in the cores might be
older, but certainly not younger, than the surrounding lower mass PMS stars.

\begin{figure}
\centering
\includegraphics[height=0.8\textheight]{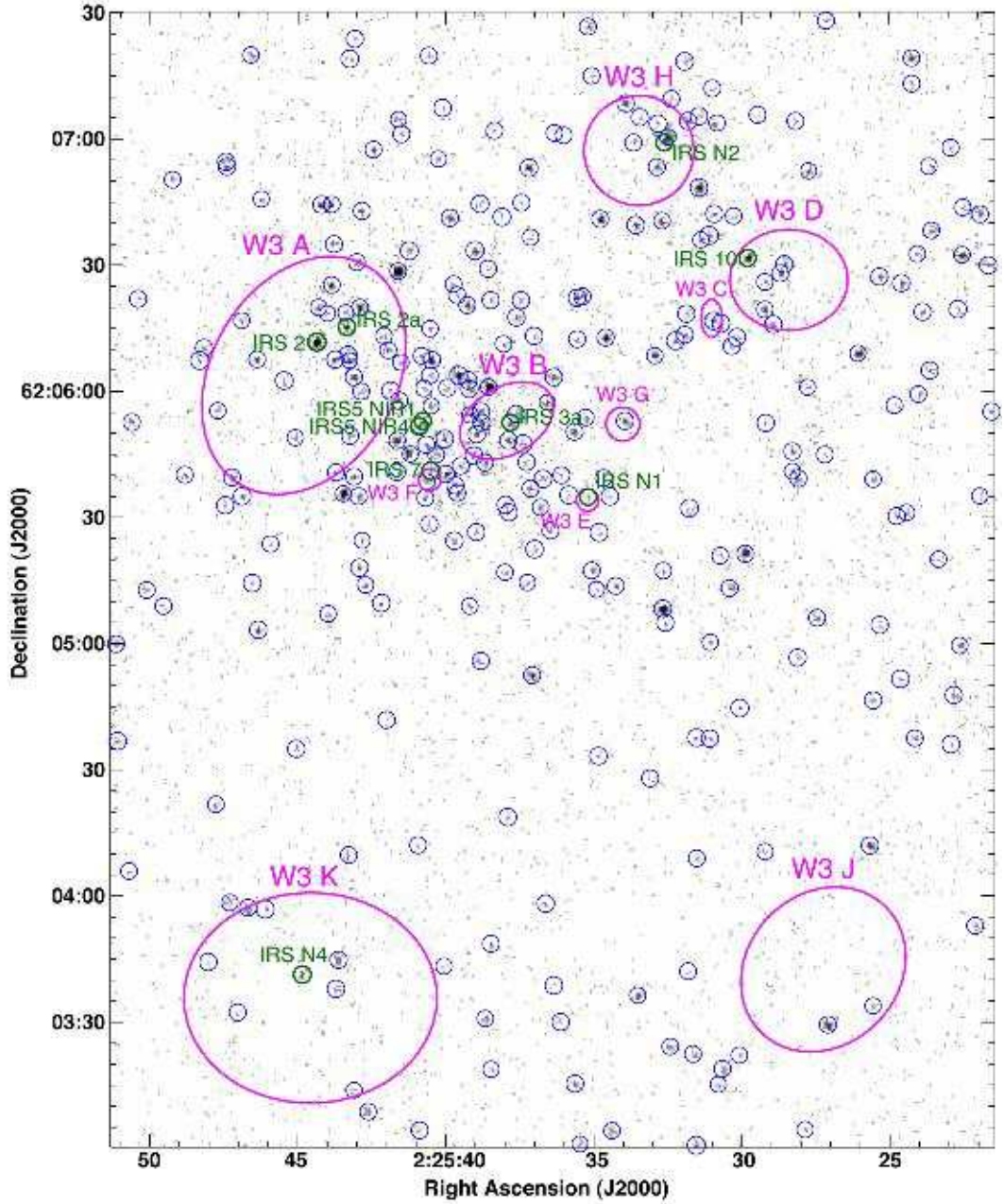}
\caption{{\it Chandra} ACIS image of W3~Main ($3.5\arcmin \times 4.5\arcmin$, 
$2.0 \times 2.6$ pc) showing individual photon events.
Well-studied radio H{\sc II} regions (magenta ellipses) and bright 
infrared sources (green circles) are labeled. \label{fig:W3Main_small}}
\end{figure}

However, several more complicated models for cluster and high-mass
star formation are consistent with our W3~Main results:
\begin{enumerate}

\item Star formation in large molecular clouds may occur inefficiently
over a prolonged period, perhaps because their dynamics are
dominated by supersonic turbulence within which only a small
fraction of the molecular material resides in dense cores at a given
moment \citep{Tan06, Krumholz07}.  The bulk of the stars may
form in a quick burst of star formation at the end of the cloud's
life, as the star formation rate becomes efficient only when
turbulence has subsided and the cloud contracts \citep{Palla00}. 
Astrophysical issues relating slow and fast star formation in clusters
are discussed by \citet{Elmegreen07}.  W3~Main exhibits a particular 
mass-dependency in its extended star formation history in that the majority of
lower mass stars appear in the widely distributed older population while
only a minority accompany the OB stars at the core.

\item The formation of massive OB stars specifically might be delayed
with respect to lower mass stars.  This delay might occur during the 
gaseous phase, where the 
formation of a high density core may be inhibited by the combined 
effect of many protostellar outflows \citep{Li04}.  Or the delay might 
occur during the stellar dynamical phase, waiting for stars to settle 
into the core gravitational potential where mergers form 
massive stars \citep{Bonnell98}.

\item Star formation may occur primarily in spatially distributed
molecular cores which, only after forming many lower mass stars over
an extended time, settle towards the cluster center where densities
are sufficiently high to form high-mass stars.  A version of this
model is described by \citet{McMillan07} as an explanation for mass
segregation in massive clusters.

\item W3~Main may contain two generations of OB stars, the latter
arising from triggering by the growing H{\sc II} regions of the
former \citep{Tieftrunk97}.  The basis for this model is the
presence of both diffuse H{\sc II} regions (W3 A, D, H, J, and K in
Figure \ref{fig:W3Main_small}) and ultracompact and hypercompact
H{\sc II} regions (W3 B, E, F, G, and M).  This would be a case of
internal triggering by W3 OB shocks rather than external triggering
by IC~1795 shocks.

\end{enumerate}

At present, we cannot differentiate between these models for
W3~Main. A useful observation would be high spatial resolution MIR
imaging to study disk properties of the lower mass {\it Chandra}
stars in the close vicinity of the OB stars (Figure
\ref{fig:W3Main_small}). This would reveal whether these 
concentrated PMS stars are younger than the more widely distributed 
PMS stars.

\section{Conclusion}

This paper introduces a new high-resolution X-ray mosaic of the W3 star
forming complex, a Large Project of the {\it Chandra X-ray Observatory}.
A rich population of $\sim 1300$ young stars is imaged and the three
well-known regions of high-mass star formation are shown to have
very different populations of low mass stars:  W3~Main is a large,
rich, nearly spherical cluster; W3(OH) lies in an elongated group of
sparse stellar clumps; and W3~North is an isolated O star without
low-mass companions. Suggestions of these differences were inferred
from earlier infrared studies, but they are more apparent here because  
the X-ray selection has the advantage of low contamination by the 
Galactic field population or diffuse interstellar emission, high 
penetration into molecular environments, and little bias towards 
stars with massive protoplanetary disks.

We emerge from this study with an improved view of star formation in 
the region. The W3(OH) structures are consistent with collect-and-collapse
triggering process caused by by shocks from the
older IC~1795 cluster,  as previously suggested. The
W3~Main cluster, however, does not show the elongated and patchy
structure of a recently triggered star cluster and appears to have
formed in an earlier episode.  Its PMS population strongly resembles
those seen in other {\em Chandra} studies of massive star-forming
regions such as those ionizing the Orion, M~17 and Rosette Nebulae.
A major difference is that the individual H{\sc II} regions in these
other clusters have already merged into a large blister and
dispersed their natal clouds.  In contrast, the W3~Main OB stars are 
very recently formed with small individual H{\sc II} regions still
embedded in a dense, clumpy molecular medium.  Star formation 
in W3 has proceeded in a prolonged fashion, and apparently with a
time-dependent Initial Mass Function.  The OB stars exciting the
hypercompact and ultracompact HII regions at the center of W3~Main
formed more recently than the hundreds of X-ray emitting
PMS stars distributed over several parsecs. W3~Main
thus becomes a critical testbed for theories of rich cluster
formation.

\acknowledgements
We thank Bruce Elmegreen (IBM) and our colleagues at Penn State --
Patrick Broos, Gordon Garmire, Kostantin Getman, Masahiro Tsujimoto,
and Junfeng Wang -- for thoughtful discussions.  Patrick Broos and
Junfeng Wang additionally provided technical assistance.  This work
was supported by the {\it Chandra} General Observer grant G05-6143X
(PI Townsley) and by the ACIS instrument team contract SV4-74018 (PI
Garmire), both issued by the {\it Chandra X-ray Observatory} Center,
which is operated by the Smithsonian Astrophysical Observatory for and
on behalf of NASA under contract NAS8-03060.

{\it Facilities:} \facility{CXO (ACIS)}

\dataset [ADS/Sa.CXO\#obs/446]  {Chandra ObsID 446}
\dataset [ADS/Sa.CXO\#obs/611]  {Chandra ObsID 611}
\dataset [ADS/Sa.CXO\#obs/5889]  {Chandra ObsID 5889}
\dataset [ADS/Sa.CXO\#obs/5890]  {Chandra ObsID 5890}
\dataset [ADS/Sa.CXO\#obs/5891]  {Chandra ObsID 5891}
\dataset [ADS/Sa.CXO\#obs/6335]  {Chandra ObsID 6335}
\dataset [ADS/Sa.CXO\#obs/6348]  {Chandra ObsID 6348}

\clearpage

% FIGURES =========================================

\clearpage

\end{document}